\documentclass[table]{ccjnl}
\usepackage{lipsum,amsmath}
\usepackage{cuted}
\usepackage{bm}
\usepackage{graphicx}
\graphicspath{{figures/}}
\usepackage{multirow}
\usepackage{xcolor}
\usepackage{subfloat}
\usepackage{xcolor}
\usepackage{xpatch}

\title{Multi-Channel Multi-Step Spectrum Prediction Using Transformer and Stacked Bi-LSTM}
\author{Guangliang Pan \inst{1}, Jie Li \inst{1,*}, Minglei Li \inst{2}}
\corinfo{lijie\_evelyn@nuaa.edu.cn}
\receiveddate{Sep. 6, 2022}
\reviseddate{Dec. 5, 2022, accepted on Aug. 3, 2023}
\Editor{}

\address[1]{College of Electronic and Information Engineering, Nanjing University of Aeronautics and Astronautics, Nanjing 211106, China}
\address[2]{College of Control Science and Engineering, China University of Petroleum (East China), Qingdao 266580, China}

\begin{document}

\maketitle

\begin{abstract}
Spectrum prediction is considered as a key technology to assist spectrum decision. Despite the great efforts that have been put on the construction of spectrum prediction, achieving accurate spectrum prediction emphasizes the need for more advanced solutions. In this paper, we propose a new multi-channel multi-step spectrum prediction method using Transformer and stacked bidirectional LSTM (Bi-LSTM), named TSB. Specifically, we use multi-head attention and stacked Bi-LSTM to build a new Transformer based on encoder-decoder architecture. The self-attention mechanism composed of multiple layers of multi-head attention can continuously attend to all positions of the multichannel spectrum sequences. The stacked Bi-LSTM can learn these focused coding features by multi-head attention layer by layer. The advantage of this fusion mode is that it can deeply capture the long-term dependence of multichannel spectrum data. We have conducted extensive experiments on a dataset generated by a real simulation platform. The results show that the proposed algorithm performs better than the baselines.
\end{abstract}
\quad

\quad

\quad
\section{Introduction}
\label{Introduction}
\subsection{Background and Motivation}
In recent years, the electromagnetic spectrum space shows a trend of integrated development of road, sea, air, sky, and network \cite{9232927, 9184022, bib3}. The influx of massive spectrum users leads to extreme shortage of spectrum resources and spectrum security problems emerge endlessly. A typical spectrum security example is that many users illegally occupy the spectrum resources of honest users (HUs) without permission. These malicious users (MUs) mainly interfere with the communication links (uplink or downlink) between honest users and base stations (H2B) and between honest users (H2H) \cite{bib4, bib5, bib6}. In this paper, we consider a strong interference scenario. Fortunately, spectrum prediction is considered as one of the key technologies to solve the above crisis \cite{bib14, bib8, bib9}. It can help users to know the future spectrum information in advance to make accurate spectrum decisions, to ensure the security of spectrum usage. Spectrum prediction has widely used in adaptive spectrum sensing \cite{bib11}, proactive spectrum mobility \cite{bib13} and dynamic spectrum access (DSA) \cite{bib10}, etc. However, for multi-channel multi-step spectrum prediction, the cumulative error will increase with the increase of the number of prediction steps \cite{shawel2019convolutional}. Moreover, existing spectrum prediction methods use recurrence and convolution structures to design algorithms, and their ability to capture long-term dependence of spectrum data is relatively limited \cite{zhou2021informer}. Moreover, multi-channel spectrum data containing multiple complex and nonlinear interference modes \cite{9761100} have weak correlation, which cannot be captured by existing algorithms. Motivated by these challenges, developing an advanced spectrum prediction algorithm has become urgent.

\subsection{Related Work}
Existing spectrum prediction methods are mainly divided into the following categories: model-driven methods and data-driven methods. Autoregressive (AR), as a classical model-driven method, is widely used in spectrum prediction. For instance, an AR-based spectrum occupancy prediction method was proposed in \cite{bib19}. In \cite{bib20}, the authors proposed an exponential moving average (EMA) based energy prediction method. However, these models rely on some ideal assumptions (e.g., linear data), while in the real world, highly nonlinear spectrum environment and endless interference make these assumptions collapse \cite{8930636}. The work of \cite{luo2021temporal} proposed a hidden Markov model (HMM) and Homotopy theory based spectrum prediction method for unmanned aerial vehicle (UAV) communications. The authors in \cite{bib16} proposed a HMM-based channel prediction method to improve the throughput for frequency-hopping cognitive radio systems. The authors in \cite{bib17} proposed a modified HMM-based single secondary-user (single-SU) prediction to minimize the negative impact of response delays caused by hardware platforms. In \cite{bib18}, a spectrum occupancy prediction method based on HMM was proposed. However, the solution of state transition probability by HMM model depends on prior information. While in practice, spectrum information is usually unavailable or imperfect. Moreover, this model requires constant training and updating of parameters, resulting in high computational costs.

Recently, data-driven DL \cite{bib21,bib22} has been regarded as a treasure by more and more researchers due to its excellent nonlinear data processing ability. In \cite{bib26,bib27}, the authors used the LSTM to implement spectrum prediction. In particular, the \cite{bib27} proposed to utilize LSTM as a deep transfer network to achieve cross-band spectrum prediction. These methods do not need prior information of the spectrum, and the prediction performance outperforms the traditional methods. However, LSTM is a unidirectional memory structure that only learns the current and previous moment information. The work of \cite{bib29} proposed a joint fine-tuned convolutional neural network (CNN) and gated recurrent unit network (GRU) learning system to achieve spectrum availability prediction. Further, authors in \cite{9625078} used LSTM, \textit{Seq-to-Seq} modeling, and attention to design a multi-channel multi-step spectrum prediction model. The work in \cite{9296309} used convLSTM networks with three components (i.e., temporal closeness, daily period, and weekly trend) to predict future spectrum situations. However, the first convolutional layer of the model will destroy the original correlation of spectrum data, which will degrade its prediction performance. These models focus on the design of complex network structures to achieve high computational complexity in exchange for performance improvement. Further, these models usually adopt recurrence and convolution structures, which are unable to attend to all positions of the spectrum data. This results in limited performance in capturing long-term dependencies. Moreover, although the proposed algorithm in \cite{9625078} incorporates LSTM and attention mechanisms, it still relies on recursive structures.   

\subsection{Contributions}
To address these challenges, we propose a novel multi-channel multi-step spectrum prediction method using Transformer and stacked bidirectional LSTM (Bi-LSTM), dispensing with recurrence and convolutions entirely, named TSB. From our knowledge, this is the first work that develops a new Transformer model to solve spectrum prediction problem by combining self-attention with stacked Bi-LSTM. In this work, we consider a scenario where a heterogeneous interfered wireless communication network and then formulate the problem of multi-channel multi-step spectrum prediction. Then, we design the TSB spectrum prediction to help users make reasonable spectrum decision. Moreover, we build a physical platform to simulate the communication interference scenario to generate a dataset to verify the reliability of the designed algorithm. Major contributions of this work are summarized below:
\begin{itemize}
	\item We propose a new Transformer (i.e., TSB) using encoder-decoder, self-attention, and stacked Bi-LSTM, which is used to achieve multi-channel multi-step spectrum prediction. 
	\item Further, TSB learns the long-term dependencies of spectrum sequences in parallel with multi-head attention, and further learns these dependencies with stacked Bi-LSTM in encoder and decoder. This fusion mode enables TSB to achieve optimal prediction performance. 
	\item We build a physical platform to perform comprehensive experiments to validate the effectiveness of the proposed algorithm under different parameter settings. The simulation results show that the proposed algorithm yields state-of-the-art prediction
	performance compared with the benchmarks.     
\end{itemize}

The remainder of this paper is organized as follows. In Section \ref{sec2}, we introduce system model and problem formulation. In Section \ref{sec3}, we introduce the proposed algorithm for multi-channel multi-step spectrum prediction. Subsequently, we discuss extensive simulation results in Section \ref{sec4}. Finally, several concluding remarks and future work are given in Section \ref{sec5}.
\begin{figure}
	\centering
	\includegraphics[width=80mm,height=64mm]{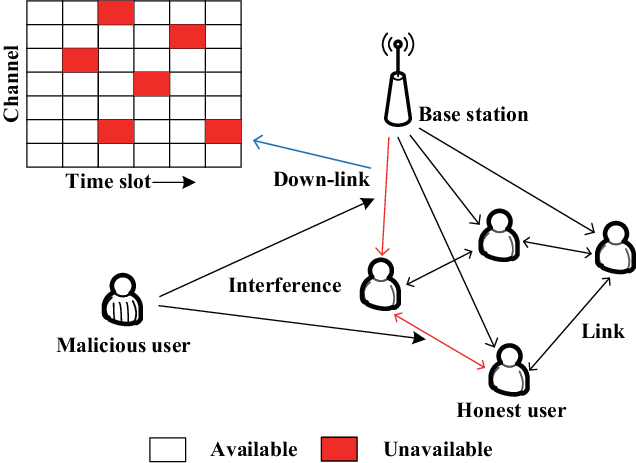}
	\caption{System model.}\label{fig1}
\end{figure}
\section{Problem Statement}
\label{sec2} 
As shown in Fig. \ref{fig1}, we consider a cognitive wireless network consisting of a MU and multiple HUs. In the system, there are several HUs communicating with the BS. Assuming that a MU interferes with the H2B downlink of the HU(s) in the $f$th channel at $t$th time slot, and the measured spectrum data can be expressed as
\begin{equation}\label{eq1}
p_{f,t} = h^z_{f,t} z_{f,t} + h^a_{f,t} a_{f,t} + d_{f,t},
\end{equation}
where $z_{f,t}$ denotes the signal strength of the HU(s), $a_{f,t}$ corresponds to the signal strength of the MU, and $d_{f,t}$ is the noise \cite{bib30}. Herein, the additive Gaussion white noise (AGWN) with mean value 0 and variance ${\sigma_d}^2$ is adopted. $h^z_{f,t}$ or $h^a_{f,t}$ is an indicator function, defined as $h^z_{f,t} = 0$ or $h^a_{f,t} = 0$ for the null hypothesis (signal is absent), and $h^z_{f,t} = 1$ or $h^a_{f,t} = 1$ for the non-null hypothesis (signal is present).  

Spectrum prediction problem is to predict future available spectrum channels based on historical spectrum data. Specifically, we first regard the measured spectrum data that changes with time under different channels as a matrix according to (\ref{eq1}). From the time-frequency interference diagram in the upper left corner of Fig. \ref{fig1}, it can be seen that the rows and columns of the matrix can be considered as radio channels and time slots respectively. Then, the measured spectrum sequence of the $f$th channel with $T$ time slots is rewritten as
\begin{equation}\label{eq2}
\bm{{\rm p}}_{f,1:T} = (p_{f,1}, \dots, p_{f,t}, \dots, p_{f,T}),
\end{equation}
where $p_{f, t}$, $f \in \{1, 2, \dots, F\}$, $t \in \{1, 2, \dots, T\}$ represents the signal power in dBm in the $f$th channel at the $t$th time slot. Further, the observed signal power matrix composed of $F$ channels with $T$ time slots is
\begin{equation}\label{eq3}
\bm{{\rm p}}_{1:F,1:T} = (\bm{{\rm p}}_{1,1:T}, \dots, \bm{{\rm p}}_{f,1:T}, \dots, \bm{{\rm p}}_{F,1:T}).
\end{equation}
Then, the multi-channel multi-step spectrum prediction problem given $F$ channels with $T$ time slots historical spectrum data $\bm{{\rm p}}_{1:F,1:T}$ can be expressed as
\begin{equation}\label{eq4}
\begin{aligned}
&\widehat{\bm{{\rm p}}}_{1:F, T+1:T+M} = \\&{\rm arg} \mathop {\rm max} \limits_{\bm{{\rm p}}_{1:F, T+1:T+M}} p(\bm{{\rm p}}_{1:F, T+1:T+M} \mid \bm{{\rm p}}_{1:F,1:T}),
\end{aligned}
\end{equation}
where $M$ stands for the prediction range. Then, based on $\widehat{\bm{{\rm p}}}_{1:F, T+1:T+M}$ in (\ref{eq4}), we consider a hard decision about the availability of the channels. We have
\begin{equation}\label{eq5}
\text{Channel state}=
\left\{
\begin{array}{rcl}
0, &\text{if}& \widehat{p}_{f,t+m} \textless \lambda\\
1, &\text{if}& \widehat{p}_{f,t+m} \geq \lambda
\end{array} \right. 
.
\end{equation} 
Here, $m=1, \dots, M$, 0/1 means the spectrum is available/unavailable, and $\lambda$ is the threshold. Finally, the spectrum availability of the $M$ time slots is predicted based on (\ref{eq5}).
\begin{figure*}
	\centering
	\includegraphics[width=0.99\textwidth]{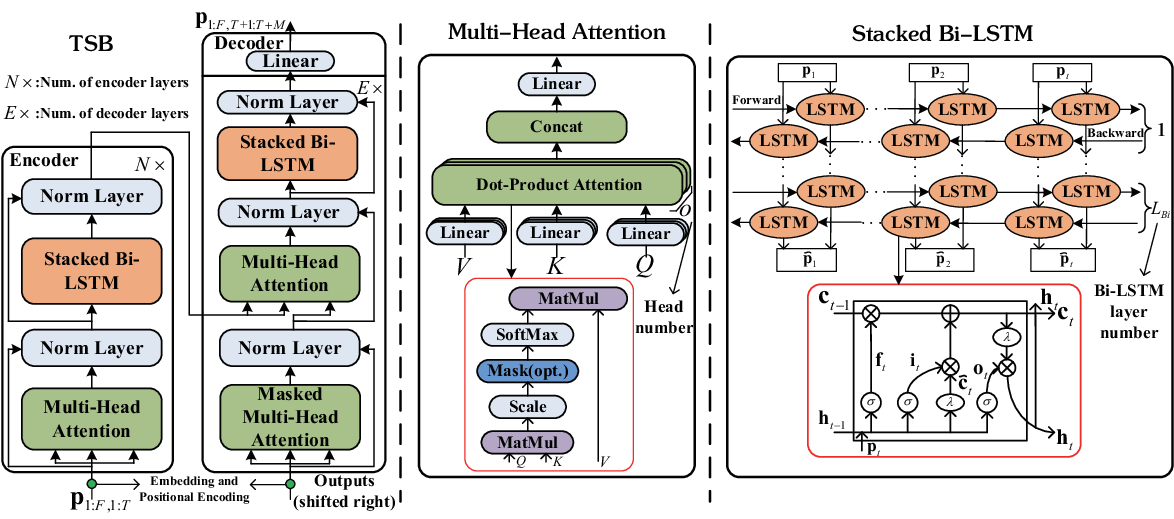}
	\caption{The architecture of the TSB algorithm (see left part).}\label{fig2}
\end{figure*}

\section{Methodology}
\label{sec3}
In this section, we introduce the proposed TSB spectrum prediction algorithm, the overall architecture of TSB is shown in Fig. \ref{fig2}. Based on Transformer, TSB includes the encoder and decoder. For the encoder, each encoder layer consists of a \textit{Multi-Head Attention} module, a \textit{Stacked Bi-LSTM} module, and a \textit{Normalized layer}. Different from the encoder, each decoder layer contains a separate linear layer with a layer normalization. Moreover, the decoder's self-attention
sub-layer uses the masking to prevent positions from attending to subsequent positions. It is worth noting that the encoder and decoder interact across information. Next, we will specifically introduce the encoder and decoder implementation process. Before introducing them, we first introduce their input. The encoder input is $\bm{{\rm p}}_{1:F,1:T}$. The decoder input is the right-shifted model output. Further, the input of the encoder and decoder is obtained by the embedding and positional encoding. In this work, these processes are the same as that of vanilla Transformer, the specific implementation process can be followed in \cite{Transformer}. 
\subsection{Encoder and Decoder}
We first define the input of the encoder as $\bm{{\rm p}}_{\text{en}}$ and the input of the decoder as $\bm{{\rm p}}_{\text{de}}$, respectively.

\textbf{Encoder:} As shown in Fig. \ref{fig2}, we assume the encoder contains $N$ encoder layers. For $n$th encoder layer, the specific implementation process can be given by
\begin{equation}\label{eq6}
\left\{
\begin{aligned}
\bm{{\rm s}}^{n,1}_{\text{en}} =&\ \small{\text{\sffamily{R-Norm}}}(\text{\sffamily{Mul-HeadAtt}}(\bm{{\rm p}}^{n-1}_{\text{en}}) + \bm{{\rm p}}^{n-1}_{\text{en}}), \\
\bm{{\rm s}}^{n,2}_{\text{en}} =&\ \small{\text{\sffamily{R-Norm}}}(\text{\sffamily{Stack-BiLSTM}}(\bm{{\rm s}}^{n,1}_{\text{en}}) + \bm{{\rm s}}^{n,1}_{\text{en}}).  
\end{aligned}
\right.
\end{equation}
The (\ref{eq6}) can be further summarized as
\begin{equation}\label{eq6sum}
\bm{{\rm p}}^{n}_{\text{en}} = \small{\text{\sffamily{Encoder}}}(\bm{{\rm p}}^{n-1}_{\text{en}}),
\end{equation}
where $\bm{{\rm p}}^{n}_{\text{en}} = \bm{{\rm s}}^{n,2}_{\text{en}}$, $n \in \{1, \dots, N\}$ denotes the output of $n$th encoder layer. When $n = 1$, $\bm{{\rm p}}^{0}_{\text{en}}$ is the embedded $\bm{{\rm p}}_{\text{en}}$. $\bm{{\rm s}}^{n,i}_{\text{en}}$, $i \in \{1,2\}$ denotes the extracted feature information after the $i$th normalization layer module in the $n$th encoder layer. $\small{\text{\sffamily{R-Norm}}}(\cdot)$ stands for a residual connection, followed by layer normalization. Further, $\small{\text{\sffamily{Mul-HeadAtt}}}(\cdot)$ and $\small{\text{\sffamily{Stack-BiLSTM}}}(\cdot)$ stand for \textit{Multi-Head Attention} module and \textit{Stacked Bi-LSTM} module, which are introduced in Sections \ref{3.1} and \ref{3.2}, respectively.

\textbf{Decoder:} As shown in Fig. \ref{fig2}, we assume the decoder contains $E$ decoder layers. For $e$th decoder layer, the specific implementation process can be given by
\begin{equation}\label{eq7}
\left\{
\begin{aligned}
\bm{{\rm s}}^{e,1}_{\text{de}}=&\ \small{\text{\sffamily{R-Norm}}}(\text{\sffamily{Mul-HeadAtt}}(\bm{{\rm p}}^{e-1}_{\text{de}}) + \bm{{\rm p}}^{e-1}_{\text{de}}), \\
\bm{{\rm s}}^{e,2}_{\text{de}}=&\ \small{\text{\sffamily{R-Norm}}}(\text{\sffamily{Mul-HeadAtt}}(\bm{{\rm s}}^{e,1}_{\text{de}}, \bm{{\rm p}}^{N}_{\text{en}}) + \bm{{\rm s}}^{e,1}_{\text{de}}), \\
\bm{{\rm s}}^{e,3}_{\text{de}}=&\ \small{\text{\sffamily{R-Norm}}}(\text{\sffamily{Stack-BiLSTM}}(\bm{{\rm s}}^{e,2}_{\text{de}}) + \bm{{\rm s}}^{e,2}_{\text{de}}).  
\end{aligned}
\right.
\end{equation}
The (\ref{eq7}) can be further summarized as
\begin{equation}\label{eq7sum}
\bm{{\rm p}}^{e}_{\text{de}} = \small{\text{\sffamily{Decoder}}}(\bm{{\rm p}}^{e-1}_{\text{de}}, \bm{{\rm p}}^{N}_{\text{en}}),
\end{equation}
where $\bm{{\rm p}}^{e}_{\text{de}} = \bm{{\rm s}}^{e,3}_{\text{de}}$, $e \in \{1, \dots, E\}$ denotes the output of $e$th decoder layer. When $e = 1$, $\bm{{\rm p}}^{0}_{\text{de}}$ is the embedded $\bm{{\rm p}}_{\text{de}}$. $\bm{{\rm s}}^{e,i}_{\text{de}}$, $i \in \{1, 2, 3\}$ denotes the extracted feature information after the $i$th $\small{\text{\sffamily{R-Norm}}}$ module in the $e$th decoder layer.  The final prediction is $\widehat{\bm{{\rm p}}}_{1:F, T+1:T+M} = \small{\text{\sffamily{Linear}}}(\small{\text{\sffamily{Norm}}}(\bm{{\rm p}}^{E}_{\text{de}}))$, where  $\small{\text{\sffamily{Linear}}}(\cdot)$ and $\small{\text{\sffamily{Norm}}}(\cdot)$ stand for the linear projection layer (which is to project $\bm{{\rm p}}^{E}_{\text{de}}$ to the target dimension) and the normalization layer, respectively. Note that the masking is used for the first multi-head attention. Then, based on these predicted values, we make a hard decision (i.e., (\ref{eq5})) to get the available spectrum resources. 
\subsection{Multi-Head Attention Module}\label{3.1}
In this subsection, we introduce the \textit{Multi-Head Attention Module} (see the \textit{middle part} in Fig. \ref{fig2}). Before introducing this module, we first introduce a single self-attention, called scaled dot-product attention. Specifically, the input of self-attention consists of queries $\bm{{\rm Q}}$ and keys $\bm{{\rm K}}$ with dimension $d_k$, and values $\bm{{\rm V}}$ with dimension $d_v$. Here, $d_k$ and $d_v$ are the dimensions of keys and values by the linear projection, respectively. The scaled dot-product attention can be given as \cite{Transformer}
\begin{equation}
\small{\text{\sffamily{Attention}}}(\bm{{\rm Q}}, \bm{{\rm K}}, \bm{{\rm V}}) = \small{\text{\sffamily{Softmax}}}(\frac{\bm{{\rm Q}}\bm{{\rm K}}^T}{d^{1/2}_k})\bm{{\rm V}},
\end{equation}
where $\small{\text{\sffamily{Softmax}}}(\cdot)$ stands for the softmax normalized function. Further, the multi-head ($o$ heads) attention can be given as
\begin{equation}
\small{\text{\sffamily{Mul-Head}}}(\bm{{\rm Q}}, \bm{{\rm K}}, \bm{{\rm V}}) = \bm{{\rm W}}_{\text{out}} \small{\text{\sffamily{Concat}}}(\text{head}_1, \cdots, \text{head}_o),
\end{equation} 
where $\text{head}_i=\small{\text{\sffamily{Attention}}}(\bm{{\rm Q}}\bm{{\rm W}}^{Q}_{i}, \bm{{\rm K}}\bm{{\rm W}}^{K}_{i}, \bm{{\rm V}}\bm{{\rm W}}^{V}_{i})$, $i \in \{1, \cdots, o\}$, and $\bm{{\rm W}}_{\text{out}} \in \mathbb{R}^{od_v \times d_{\text{model}}}$ ($d_{\text{model}}$ is the dimension of model input) is the projection of final linear layer. Here, $\bm{{\rm W}}^{Q}_{i} \in \mathbb{R}^{d_{\text{model}} \times d_k}$, $\bm{{\rm W}}^{K}_{i} \in \mathbb{R}^{d_{\text{model}} \times d_k}$, and $\bm{{\rm W}}^{V}_{i} \in \mathbb{R}^{d_{\text{model}} \times d_v}$ are the projections by the linear layer, respectively. This self-attention can attend to all positions of spectrum data and can learn long-range dependencies. Further, multi-head attention is calculated in parallel mode to reduce time consumption.
 
\subsection{Stacked Bi-LSTM Module}\label{3.2}
We design a \textit{stacked Bi-LSTM} module (see the \textit{right part} in Fig. \ref{fig2}), which is composed of multiple Bi-LSTM layers. Each Bi-LSTM layer contains multiple LSTMs connected by bidirectional rules. Different from vanilla LSTM, Bi-LSTM combines forward and backward propagation. Taking the stacked Bi-LSTM module of $n$th encoder layer for an exmple, according to (\ref{eq6}), the input is the $\bm{{\rm s}}^{n,1}_{\text{en}}$. For a single LSTM, which consists of a forgetting gate $\bm{{\rm f}}_t$, an input gate $\bm{{\rm i}}_t$ and an output gate $\bm{{\rm o}}_t$. The $\bm{{\rm f}}_t$ consists of previous layer out $\bm{{\rm h}}_{t-1}$ and current input $\bm{{\rm p}}_t$ (i.e., $\bm{{\rm s}}^{n,1}_{\text{en}}$) with weights $\bm{{\rm W}}_f, \tilde{\bm{{\rm W}}}_f$ and bias $\bm{{\rm b}}_f$,  which can be given as\cite{9743298} 
\begin{equation}\label{eq12}
\bm{{\rm f}}_t=\sigma(\bm{{\rm W}}_f \bm{{\rm p}}_t + \tilde{\bm{{\rm W}}}_f \bm{{\rm h}}_{t-1}+\bm{{\rm b}}_f),
\end{equation}
where $\sigma$ is sigmoid active funcation. Then, the spectrum sequence is fed into $\bm{{\rm i}}_t$, which can be given as
\begin{equation}\label{eq13}
\bm{{\rm i}}_t=\sigma(\bm{{\rm W}}_i \bm{{\rm p}}_t + \tilde{\bm{{\rm W}}}_i \bm{{\rm h}}_{t-1}+\bm{{\rm b}}_i),
\end{equation}
where $\bm{{\rm W}}_i$, $\tilde{\bm{{\rm W}}}_i$, and $\bm{{\rm b}}_i$ are the weights and bias of the input gate, respectively. Here, LSTM has a status update unit, can be given by
\begin{equation}\label{eq14}
\tilde{\bm{{\rm c}}}_t=\lambda(\bm{ {\rm W}}_c \bm{{\rm p}}_t  + \tilde{\bm{ {\rm W}}}_c \bm{{\rm h}}_{t-1} +\bm{{\rm b}}_c),
\end{equation}
where $\lambda$ is tanh active funcation, $\bm{{\rm W}}_c$, $\tilde{\bm{ {\rm W}}}_c$, and $\bm{{\rm b}}_c$ are the weights and bias of update unit, respectively. Then, the data is updated by
\begin{equation}\label{eq15}
\bm{{\rm c}}_t=\bm{{\rm f}}_t \odot \bm{{\rm c}}_{t-1}+\bm{{\rm i}}_t \odot \tilde{\bm{{\rm c}}}_t,
\end{equation}
where $\odot$ stands for the operation of element-wise multiplication. Then, the updated data enters $\bm{{\rm o}}_t$,
\begin{equation}\label{eq16}
\bm{{\rm o}}_t=\sigma(\bm{{\rm W}}_o \bm{{\rm p}}_t + \tilde{\bm{{\rm W}}}_o \bm{{\rm h}}_{t-1} + \bm{{\rm b}}_o),
\end{equation}
\begin{equation}\label{eq161}
\bm{{ \rm h}}_t = \bm{{\rm o}}_t \odot \lambda(\bm{{\rm c}}_t),
\end{equation}
where $\bm{{ \rm h}}_t$ is the final output. $\bm{{\rm W}}_o$, $\tilde{\bm{{\rm W}}}_o$, and $\bm{{\rm b}}_o$ are the weights and biase of the output gate, respectively.

For the training of a single Bi-LSTM layer, the forward output $\bm{{\rm h}}^{\text{cell,f}}_t$ and backward output $\bm{{\rm h}}^{\text{cell,b}}_t$ are first calculated by (\ref{eq161}). Then, the output of the Bi-LSTM layer is given by
\begin{equation}\label{26}
\bm{{\rm h}}^{\text{cell,Bi}}_t = \sigma_{\text{f, b}}(\bm{{\rm h}}^{\text{cell,f}}_t, \bm{{\rm h}}^{\text{cell,b}}_t),
\end{equation}
where the $\sigma_{\text{f, b}}$ function stands for the concatenate merge mode. Futher, the output of $L_{\text{Bi}}$th Bi-LSTM layer is given by
\begin{equation}\label{27}
\bm{{\rm h}}^{\text{cell,Bi}}_{t,L_{\text{Bi}}} = \sigma_{\text{f, b}}(\bm{{\rm h}}^{\text{cell,f}}_{t,L_{\text{Bi}}}, \bm{{\rm h}}^{\text{cell,b}}_{t,L_{\text{Bi}}}).
\end{equation}
	
The stacked Bi-LSTM module can extract the features of spectrum data layer by layer. Different from the traditional fully connected layer in the vanilla Transformer, this design can further improve the prediction performance of the proposed algorithm.
 
\subsection{Train Rules and Online Prediction}
Before online spectrum prediction, we need to train the proposed algorithm with train rules. We first adopts the mean squared error (MSE) with  $L_2$ regularization loss function to improve training speed, which is
\begin{equation}\label{loss}
\psi = \underbrace{\parallel \bm{{\rm \hat{p}}}_{t} - \bm{{\rm p}}_{t} \parallel^{2}_{2}}_{\text{MSE}} + \underbrace{\eta \dfrac{1}{2}( \sum_{n=1}^{N} (\bm{{\rm W_{en}}}^{n})^2 + \sum_{e=1}^{E} (\bm{{\rm W_{de}}}^{e})^2)}_{L_2 \text{regularization}},
\end{equation}
where $\widehat{\bm{{\rm p}}}_t$ and $\bm{{\rm p}}_t$ are the predicted values and real values at time $t$, $t \in \{1, \cdots, M\}$, respectively, $\eta$ is $L_2$ regularization coefficient, $\bm{{\rm W_{en}}}^{n}$ is the wight of $n$th encoder layer, and $\bm{{\rm W_{de}}}^{e}$ is the weight of $e$th decoder layer. Let $\bm{{\rm U}} = \{\bm{{\rm W_{en}}}^{n}, \bm{{\rm W_{de}}}^{e}, \bm{{\rm b}}\}$, where $\bm{{\rm b}}$ is the bias of the proposed algorithm. Further, the objective function can be given by
\begin{equation}
\bm{{\rm U}}^{*} = {\rm arg} \mathop {\rm min} \limits_{\bm{{\rm U}}} \psi.
\end{equation}
To get the optimal $\bm{{\rm U}}^{*}$, we use the gradient descent algorithm \cite{bib32,bib33}:
\begin{equation}\begin{split}\label{eq25}
\bm{{\rm W_{en}}}^{n}(s+1) \leftarrow& \bm{{\rm W_{en}}}^{n}(s)-\alpha \dfrac{\partial \psi(\bm{{\rm p}}_t,\bm{{\rm \hat{p}}}_{t})}{\bm{{\rm W_{en}}}^{n}},\\
\bm{{\rm W_{de}}}^{e}(s+1) \leftarrow& \bm{{\rm W_{de}}}^{e}(s)-\alpha \dfrac{\partial \psi(\bm{{\rm p}}_t,\bm{{\rm \hat{p}}}_{t})}{\bm{{\rm W_{de}}}^{e}},\\
\bm{{\rm b}}(s+1) \leftarrow& \bm{{\rm b}}(s)-\alpha \dfrac{\partial \psi(\bm{{\rm p}}_t,\bm{{\rm \hat{p}}}_{t})}{\bm{{\rm b}}},
\end{split}\end{equation}
where $\bm{{\rm W_{en}}}^{n}(s)$, $\bm{{\rm W_{de}}}^{e}(s)$, and $\bm{{\rm b}}(s)$ are weights and bias of $s$th training step. $\alpha$ is the learning rate. 

We train TSB according to the above training rules to achieve optimal prediction performance.

{\tabcolsep=4pt\small 
	\begin{center}
		\begin{tabular}{lp{75mm}} \hline
			\multicolumn{2}{l}{{\bf Algorithm 1}~~ The proposed TSB algorithm}\\ \hline
			&\textbf{Input}: $\bm{{\rm p}}_{1:F,1:T}$\\
			&\textbf{Output}:$\widehat{\bm{{\rm p}}}_{1:F, T+1:T+M}$\\
			&\textbf{Initialization}: Encoder and decoder layers $N$, $E$, training steps $S_{\text{train}}$, learning rate $\alpha$, $L_2$ regularization coefficient $\eta$, randomize initial weights and bias $\bm{{\rm U}}$;\\
			&\textbf{1) Off-line training}:\\	
			&\textbf{For} $1 \leq t \leq S_{\text{train}}$ \\
			&\quad Based on (\ref{eq6}) $\sim$ (\ref{eq7sum}) calculate the output of model;\\
			&\quad Based on (\ref{loss}) calculate the loss:\\
			&\quad $\psi = \parallel \bm{{\rm \hat{p}}}_{t} - \bm{{\rm p}}_{t} \parallel^{2}_{2} + \eta \dfrac{1}{2}( \sum_{n}^{N} (\bm{{\rm W_{en}}}^{n})^2 + \sum_{e}^{E}(\bm{{\rm W_{de}}}^{e})^2)$ \\
			&\quad Based on (\ref{eq25}) update $\bm{{\rm U}}$.\\
			&\textbf{end}\\
			&\textbf{2) Online prediction}:\\
			&\quad Based on (\ref{eq26}) complete online prediction:\\
			&\quad $\widehat{\bm{{\rm p}}}_{1:F, T+1:T+M} = \text{TSB}_{\text{Trained}}(\bm{{\rm p}}_{1:F,1:T})$ \\
			&\quad Make a spectrum decision by a hard decision.\\
			\hline
		\end{tabular}
\end{center}}

Then, the optimized TSB achieves online spectrum prediction, which is given by
\begin{equation}\label{eq26}
\widehat{\bm{{\rm p}}}_{1:F, T+1:T+M} = \text{TSB}_{\text{Trained}}(\bm{{\rm p}}_{1:F,1:T}).
\end{equation}
Finally, we make a spectrum decision by a hard decision.

We verify the performance of the proposed algorithm on a test set generated by a real simulation platform. The error between the prediction and the real values can be expressed as
\begin{equation}
\bm{{\rm\theta}} = \widehat{\bm{{\rm p}}}_{1:F, T+1:T+M}-\bm{{\rm p}}_{1:F, T+1:T+M},
\end{equation}
the prediction accuracy of the proposed algorithm is calculated by comparing the error $\bm{{\rm\theta}}$ with the threshold. The detail of the achieving process of the proposed algorithm can be found in Algorithm 1.

\section{Simulation results and analysis}\label{sec4}
\subsection{Settings}
\begin{figure}
	\centering
	\includegraphics[width=78mm,height=56mm]{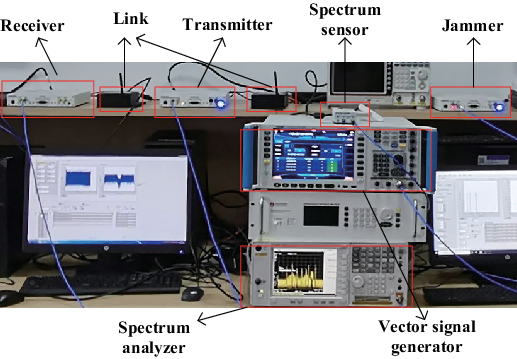}
	\caption{Simulation platform.}\label{fig3}
\end{figure}

\textit{1) Dataset:} The real experiment platform is shown in Fig. \ref{fig3}. We simulate a MU to interfere with the wireless communication link. The jammer uses USRP 2943R. Multiple interference modes are adopted (e.g., sweep interference, fixed frequency interference, etc.), the interference to signal ratio is 10 dB, and interference power exceeds -50 dBm. Spectrum sensor monitoring range is 50 MHz to 2.2 GHz. Transmitter and receiver operating power is less than -50 dBm. The spectrum analyzer is used to sense spectrum data. Herein, either interfering H2B or interfering H2H is considered interfering. Due to the laboratory scenario, the distance between the jammer antenna and the user antenna is relatively close, so channel fading is not considered. The HU nodes are assumed to communicate between 1300 MHz and 1700 MHz, and every 0.4 MHz is divided into a communication channel. Further, twenty time slots are considered as an interference period. The power is sensed every 0.1 seconds. If the measured power exceeds -50 dBm, the channel cannot be communicated at the current time. We use the $K$-fold ($K=5$) to split the spectrum data into the training set, validation set, and test set in a 5:1:1 ratio.

\textit{2) Implementation details:} We run all the experiments on a PC with 3.00 GHz Intel Core i7-9700 CPU, NVIDIA Quadro P2200 graphic, and 16 GB RAM using the Pytorch 1.7.1 with python language. The TSB adopts 3 encoder layers (i.e., $N = 3$) and 3 decoder layers (i.e., $E = 3$). The loss function is the root mean square error (MSE). The number of train epochs is set as 20. The batch size is 32. The early stop training method is used for our model, and the value of patience is 6. Adam optimizer is used to improve the convergence rate of the model. The initial learning rate is 0.001. The number of heads is set as 8.

\textit{3) Baselines and evaluation metric:} We compare the proposed algorithm with the following three baseline methods:

\begin{itemize}
\item \textbf{DCG}: A deep neural network based spectrum availability prediction model proposed in \cite{bib29}, which is a joint CNN and GRU network. 
\item \textbf{LSTM-Attention}: A special kind of RNN, which is used for multi-channel multi-step spectrum prediction in combination with attention mechanism and \textit{Seq-to-Seq} modeling \cite{9625078}.
\item \textbf{ConvLSTM}: An LSTM network with the convolution structure, which is used for spectrum prediction in \cite{9296309}.
\end{itemize}

We use the root mean squared error (RMSE) in dB to quantify the (relative) magnitude of the prediction error \cite{bib30}.

\subsection{Numerical Analysis}\label{quan}
Under SNR = 40 dB with 1304 MHz, we evaluate the performance of the proposed TSB by conducting experiments with a spectrum dataset.  

\begin{figure*}
	\centering
	\subfloat[TSB (ours), SNR = 40 dB]{
		\centering
		\includegraphics[width=72mm,height=25mm]{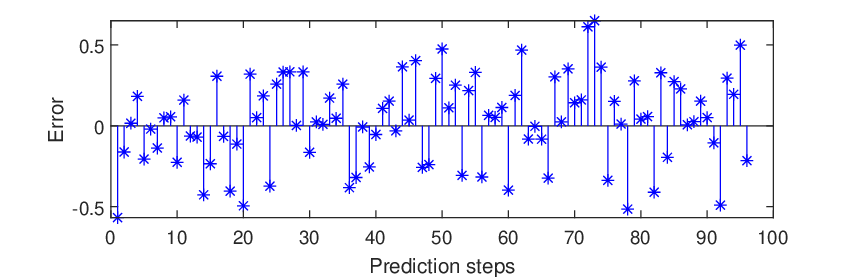}
		\label{fig6a}}%
	\subfloat[ConvLSTM, SNR = 40 dB]{
		\centering
		\includegraphics[width=72mm,height=25mm]{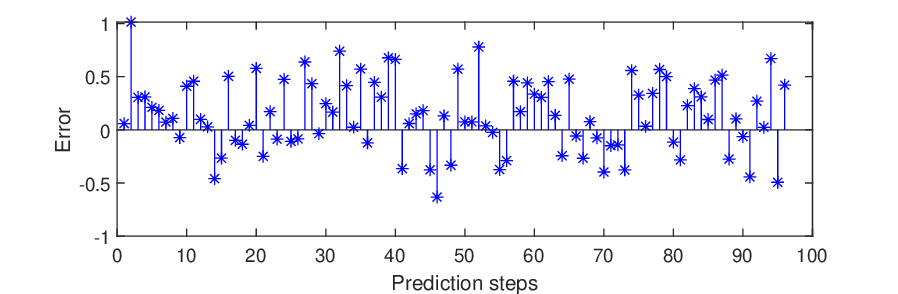}
		\label{fig6b}}%
	\\
	\subfloat[LSTM-Attention, SNR = 40 dB]{
		\centering
		\includegraphics[width=72mm,height=25mm]{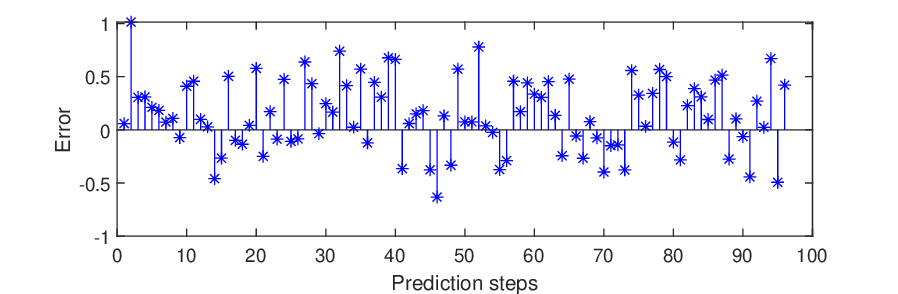}
		\label{fig6c}}%
	\subfloat[DCG, SNR = 40 dB]{
		\centering
		\includegraphics[width=72mm,height=25mm]{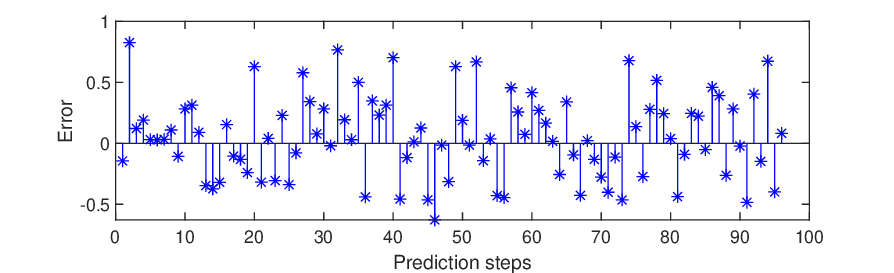}
		\label{fig6d}}%
	\caption{The prediction results comparison of the proposed TSB with three benchmarks.}\label{fig6}
\end{figure*}

As shown in Fig. \ref{fig6}, we can see that the prediction accuracy of the proposed TSB is obviously higher than three benchmarks. Specifically, LSTM with attention has the worst prediction performance. Further, the prediction error of the three benchmarks has a great fluctuation. Compared with three benchmarks, the proposed TSB has a minimum prediction error. For instance, we take a normalized prediction error of 0.5 as a threshold. There are only 5 time slots in which the prediction error of the proposed TSB exceeds 0.5. In contrast, ConvLSTM, LSTM with attention, and DCG have 14, 17, and 13 time slots, respectively. This indicates that the ability of the three baselines to capture interference correlation information is limited. Further, the proposed TSB uses multi-head self-attention mechanism, dispensing with recurrence and convolutions entirely. 
\begin{figure}%
	\centering
	\includegraphics[width=72mm,height=63mm]{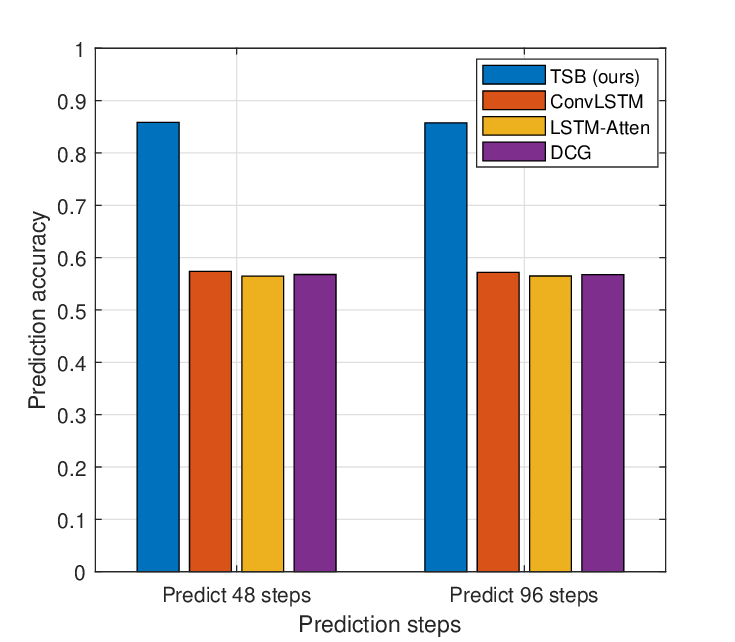}
	\caption{The prediction accuracy comparison of the proposed TSB with three benchmarks.}\label{fig7}
\end{figure}
\begin{figure}[h]%
	\centering
	\includegraphics[width=72mm,height=63mm]{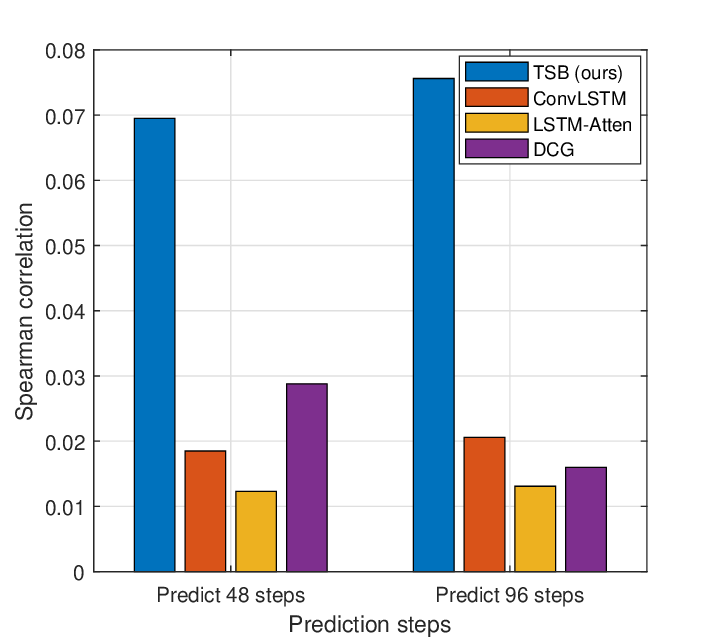}
	\caption{The Spearman correlation comparison of the proposed TSB with three benchmarks.}\label{fig8}
\end{figure}

Fig. \ref{fig7} shows that the prediction accuracy of the proposed TSB is better than three benchmarks with predicting future 48 steps and 96 steps under SNR = 40 dB. Specifically, for prediction with 48 steps, the prediction accuracy of our TSB increases by \textbf{49.65} \% compared to ConvLSTM, \textbf{52.06} \% compared to LSTM with attention, and \textbf{51.18} \% compared to DCG. For prediction with 96 steps, the prediction accuracy of our TSB increases by \textbf{49.94} \% compared to ConvLSTM, \textbf{51.80} \% compared to LSTM with attention, and \textbf{51.13} \% compared to DCG. Then, we use the Spearman's correlation coefficient \cite{bib39} to analyze the prediction performance of the proposed TSB. The rank correlation coefficient between predicted results and real values can be given by
\begin{equation}
\kappa = 1-\dfrac{6 \sum_{k=1}^{M} {d_k}^2}{M(M^2-1)}, k = 1, 2, ..., M,
\end{equation}
where $d_k$ represents the rank difference between the predicted values and the real values of the $k$th time slot. The values of $\kappa$ range from -1 to 1, with extreme values indicating (inversely) fully correlation. In general, the correlation of predicted and real sequences is higher as value of the coefficient $\kappa$ is closer to 1. Herein, $M = \{48, 96\}$.

Fig. \ref{fig8} shows the $\kappa$ on the four methods with different prediction steps in test data. From Fig. \ref{fig8}, we see the $\kappa$ of the proposed TSB is higher than 0.03. While the average $\kappa$ of the three benchmarks are less than 0.03. For instance, when the number of predicted steps is 48, the $\kappa$ of the proposed TSB outperforms ConvLSTM, LSTM with attention, and DCG by 0.0510, 0.0572, and 0.0407, respectively. This is because $\kappa$ represents the degree of correlation and has no direct relationship with the calculation of predicted accuracy. Note that the coefficient $\kappa$ is a necessary rather than sufficient condition for prediction accuracy. Moreover, we note that the spectrum data presents a weak correlation, but the TSB still maintains a high prediction accuracy.
\begin{figure}[h]%
	\centering
	\includegraphics[width=77mm,height=66mm]{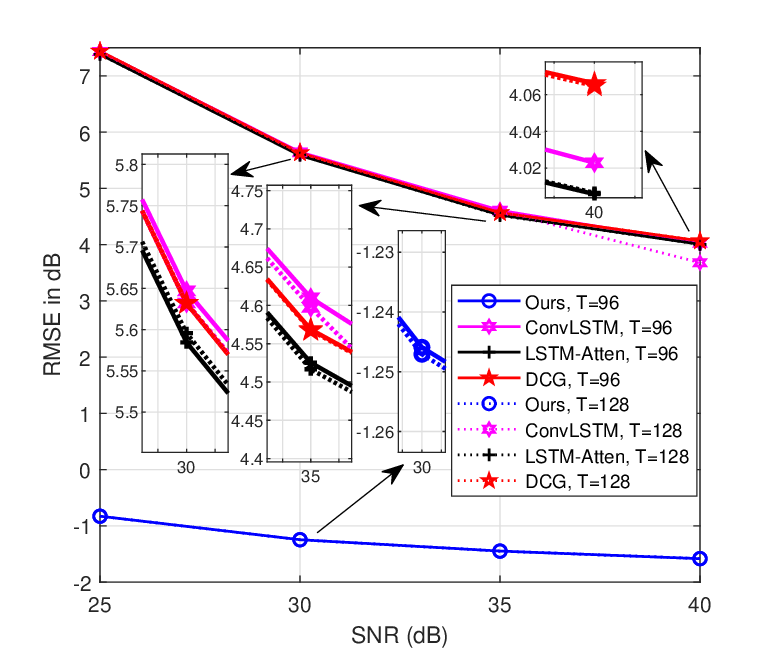}
	\caption{Comparison of the proposed TSB with three benchmarks under different input lengths.}\label{fig9}
\end{figure}
\begin{figure}[h]%
	\centering
	\includegraphics[width=77mm,height=66mm]{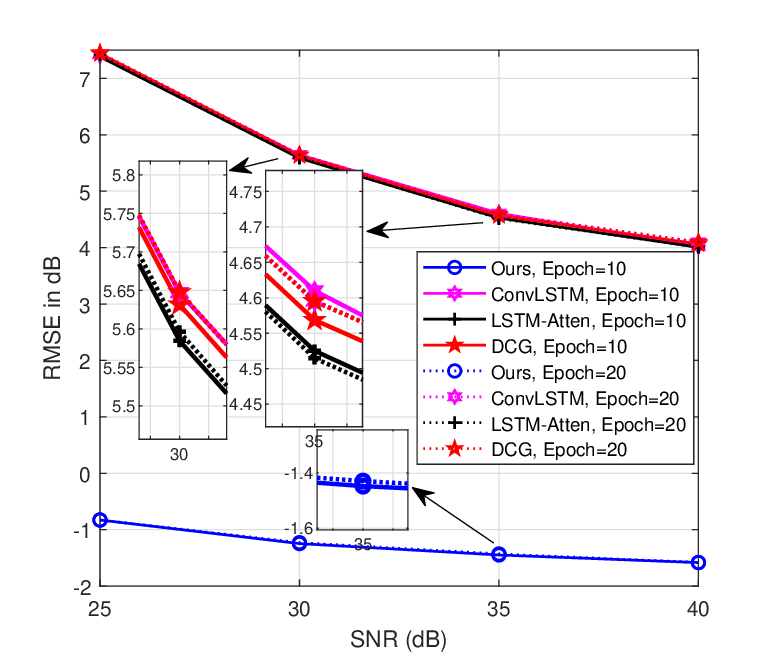}
	\caption{Comparison of the proposed TSB with three benchmarks under different epochs.}\label{fig10}
\end{figure}
\begin{figure}[h]%
	\centering
	\includegraphics[width=78mm,height=66mm]{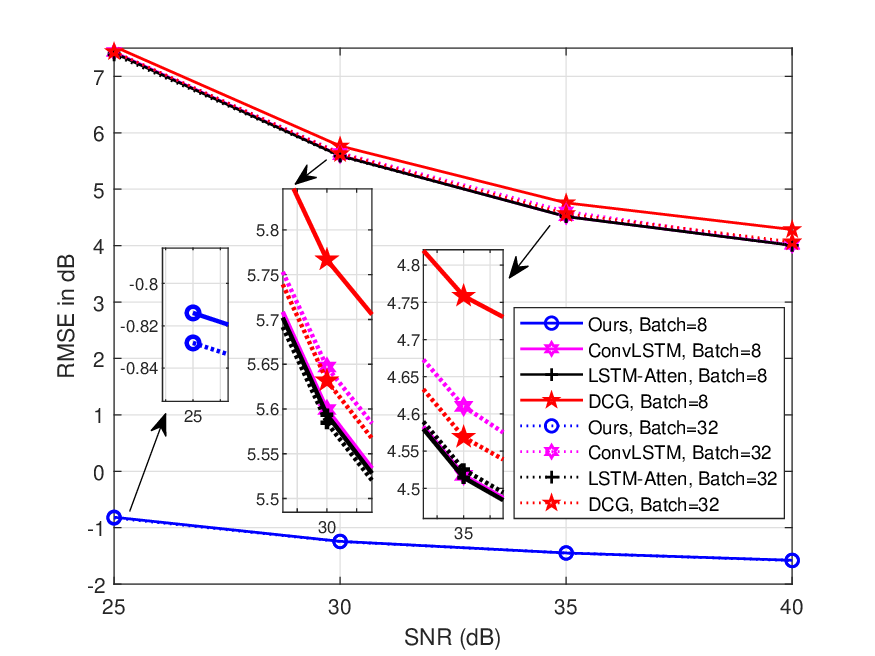}
	\caption{Comparison of the proposed TSB with three benchmarks under different batchs.}\label{fig11}
\end{figure}
\begin{figure}[h]%
	\centering
	\includegraphics[width=77mm,height=66mm]{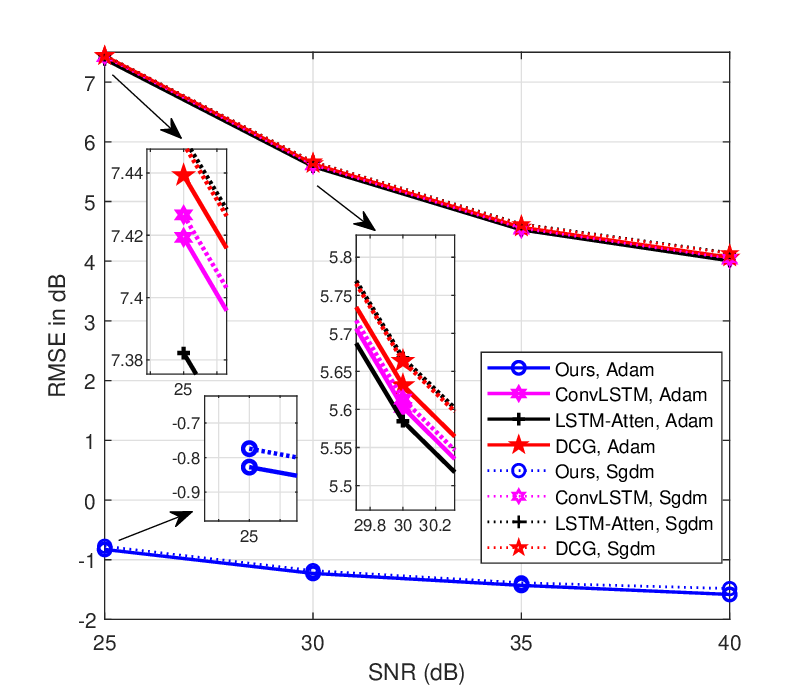}
	\caption{Comparison of the proposed TSB with three benchmarks under different gradient descent methods.}\label{fig12}
\end{figure} 
\subsection{Impact of Input Length and Epoch}
In Fig. \ref{fig9}, we consider the effect of the input length on the prediction performance of the four algorithms. Compared with convLSTM, the RMSE of the proposed TSB does not change significantly with the input length increasing from 96 steps to 128 steps, indicating that TSB is robust to the input length. Moreover, taking the input-96-predict-48 with SNR = 25 dB for an example, TSB has a RMSE decrease of \textbf{85.07\%} (5.5361 $\rightarrow$ 0.8264) to convLSTM, \textbf{84.90\%} (5.4729 $\rightarrow$ 0.8264) to LSTM with attention, and \textbf{85.10\%} (5.5453 $\rightarrow$ 0.8264) to DCG. Fig. \ref{fig10} shows the effect of the epoch. Specifically, taking the input-96-predict-48 with 20 epochs under SNR = 30 dB for an example, TSB has a RMSE decrease of \textbf{79.55\%} (3.6702 $\rightarrow$ 0.7506) to convLSTM, \textbf{79.25\%} (3.6179 $\rightarrow$ 0.7506) to LSTM with attention, and \textbf{79.48\%} (3.6573 $\rightarrow$ 0.7506) to DCG. Overall, the prediction performance of the proposed algorithm is significantly better than that of the three baselines under different SNRs is robust to input length and the number of epoch.
\subsection{Impact of Batch and Optimizer}
Fig. \ref{fig11} shows the effect of the batch. Compared with DCG, the RMSE of the proposed TSB does not change significantly with the batch size decreases from 32 to 8, indicating that TSB is robust to the batch size. Moreover, for input-96-predict-48, taking the batch size with 8 under SNR = 30 dB for an example, TSB has a RMSE decrease of \textbf{79.30\%} (3.6303 $\rightarrow$ 0.7513) to convLSTM, \textbf{79.28\%} (3.6254 $\rightarrow$ 0.7513) to LSTM with attention, and \textbf{80.09\%} (3.7729 $\rightarrow$ 0.7513) to DCG. From Fig. \ref{fig12}, we consider analyzing the prediction performance of the proposed TSB in different optimizers (i.e., adam \cite{bib33} and sgdm \cite{bib40}). Specifically, taking the adam optimizer under SNR = 35 dB for an example, TSB has a RMSE decrease of \textbf{74.64\%} (2.8376 $\rightarrow$ 0.7195) to convLSTM, \textbf{74.62\%} (2.8348 $\rightarrow$ 0.7195) to LSTM with attention, and \textbf{74.87\%} (2.8633 $\rightarrow$ 0.7195) to DCG. Overall, the prediction performance of the proposed TSB is significantly better than that of the three baselines under different SNRs is robust to the batch and optimizer.
\begin{table}
	\begin{center}		
		\caption{Ablation study on the choices of the number of encoder and decoder layers}\label{tab7}
		\begin{tabular}{c c c c c}
			\hline  
			Number of layers & $\{2, 2\}$ & $\{3, 3\}$ & $\{4, 4\}$\\
			\midrule
			RMSE (48 steps)  & 0.6849 & \textbf{0.6838} & 0.6842\\  
			RMSE (96 steps)  & 0.6847 & \textbf{0.6845} & 0.6851\\
			\hline
		\end{tabular}		
	\end{center}
\end{table}
\begin{table}
	\begin{center}		
		\caption{Ablation study on the choices of the number of heads}\label{tab8}
		\begin{tabular}{c c c c}
			\hline  
			Number of heads& 8 & 10 & 12 \\
			\midrule
			RMSE (48 steps) & \textbf{0.6838} & 0.6840 & 0.6848 \\  
			RMSE (96 steps) & 0.6845 & \textbf{0.6843} & 0.6845 \\
			\hline
		\end{tabular}		
	\end{center}
\end{table}
\subsection{Ablation Studies}
In this subsection, we perform ablation studies to analyze the efficacy of each design choice in the proposed algorithm. The input length is set to 96. Due to the limited computational resources, we train each model for 20 epochs, while other training settings keep the same as in Section 4.1.

\textbf{Number of encoder and decoder layers.} We perform ablation studies of the number of encoder and decoder layers. Here, we set the number of encoder and decoder layers to $\{2, 2\}$, $\{3, 3\}$, and $\{4, 4\}$, the results are shown in Table \ref{tab7}. From Table \ref{tab7}, for 48 and 96 steps prediction, we see that TSB achieves an optimal prediction performance under $\{3, 3\}$ settings. Thus, we apply $\{3, 3\}$ as the default hidden units choice.

\textbf{Number of heads choices.} We conduct experiments for different head numbers $\{8, 10, 12\}$. As shown in Table \ref{tab8}, for 48 steps prediction, we see that TSB achieves the smallest RMSE with the number of heads set to 8. While for 96 steps prediction, TSB achieves optimal prediction performance under a 10-head setting. Increasing the number of heads increases the computational memory. Thus, we use a 8-head as the base configuration.

\textbf{Number of stacked Bi-LSTM choices.} To select the optimal Bi-LSTM layers, we set the number of Bi-LSTM layers to $\{1, 2, 3\}$. For 48 and 96 steps prediction, as shown in Table \ref{tab9}, when the number of Bi-LSTM layers is $2$, RMSE is minimum. The RMSE deteriorates rapidly when the number of Bi-LSTM layers is $1$. Compared with $2$ Bi-LSTM layers, although the RMSE does not change significantly when the number of Bi-LSTM layers is $3$, a large number of Bi-LSTM layers will bring extra time cost of training model. Thus, the number of Bi-LSTM layers is set to $2$.

\textbf{Learning rate choices.} We perform ablation studies of the learning rate. Here, we try to set the learning rate to $\{0.01, 0.001, 0.0001\}$, the results are shown in Table \ref{tab10}. For the proposed algorithm with 48 and 96 steps prediction, when the learning rate is 0.001, TSB has an optimal prediction performance. Hence, the learning rate is set to 0.001.

\begin{table}
	\begin{center}		
		\caption{Ablation study on the choices of the number of Bi-LSTM layers}\label{tab9}
		\begin{tabular}{c c c c}
			\hline  
			Number of Bi-LSTMs & 1 & 2 & 3 \\
			\midrule
			RMSE (48 steps) & 0.6851 & \textbf{0.6844} & 0.6844 \\  
            RMSE (96 steps) & 0.6855 & \textbf{0.6845} & 0.6850 \\  
			\hline
		\end{tabular}		
	\end{center}
\end{table}
\begin{table}
	\begin{center}		
		\caption{Ablation study on the choices of learning rate}\label{tab10}
		\begin{tabular}{c c c c}
			\hline  
			Learning rate & 0.01 & 0.001 & 0.0001 \\
			\midrule
			RMSE (48 steps) & 0.6840 & \textbf{0.6835} & 0.6838 \\
			RMSE (96 steps) & 0.6880 & \textbf{0.6829} & 0.6844 \\  
			\hline
		\end{tabular}		
	\end{center}
\end{table}

\textbf{Other design choices.} To validate the effectiveness of other design choices like the number of LSTM hiddens and loss function, we add these components one by one to the proposed algorithm. According to the above comparison criteria, the optimal choices of other designs are shown in Section 4.1.
\section{Conclusion and Future Work}\label{sec5}
In this paper, we have proposed a TSB method for multi-channel multi-step spectrum prediction. Specifically, a communication link interference scenario is considered to model the multi-channel multi-step spectrum prediction problem. Then, we design a prediction model composed of the encoder-decoder architecture, multi-head attention, and stacked Bi-LSTM, which can accurately capture the long-term temporal correlation of spectrum data. On a real experimental platform, simulation results have shown that the proposed prediction algorithm dramatically outperforms different benchmarks under different parameter settings.

However, with the number of encoder and decoder layers increasing, the training time and memory resources of the proposed method rise dramatically. The tradeoff between prediction performance and training cost needs to be further explored. On the other hand, in dynamic wireless communication systems, the distribution of spectrum data may change, and the proposed long-term prediction model will fail in this case. Transfer learning is a promising tool for migrating models across different scenarios \cite{9664805, 9352556}. Future work will consider applying transfer learning to the proposed model. The transfer learning module tunes the existing model’s parameters accordingly when the distribution of data is changed.

\bibliographystyle{IEEEtran}
\bibliography{myref}
\section*{Biographies}
\begin{CCJNLbiography}{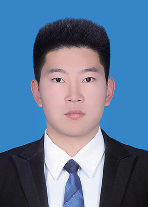}{Guangliang Pan}
received the B.Eng. degree in automation and the M.Eng. degree in control engineering from Qilu University of Technology (Shandong Academy of Sciences), Jinan, China, in 2017 and 2020, respectively. He is currently pursuing the Ph.D. degree with the College of Electronic and Information Engineering, Nanjing University of Aeronautics and Astronautics. He is also studying as a Visiting Ph. D. Student with the Pillar of Information Systems Technology and Design, Singapore University of Technology and Design. His current research interests include cognitive radio and deep learning.

\end{CCJNLbiography}
\begin{CCJNLbiography}{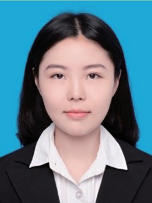}{Jie Li}
received the B.S. degree in electrical engineering and the Ph. D. degree in signal and information processing from Xidian University, Xi’an, China, in 2013 and 2020, respectively. She was studying as a Visiting Ph. D. Student with the Department of Electrical and System Engineering, Washington University in St. Louis, St. Louis, MO, USA, from Oct. 2017 to Oct. 2019. Since 2021, she has been an associate Professor in the College of Electronic and Information Engineering, Nanjing University of Aeronautics and Astronautics, Nanjing, China. Her current research interests include dynamic cognitive system, adaptive array signal processing, MIMO radar signal processing, and image processing.
\end{CCJNLbiography}

\begin{CCJNLbiography}{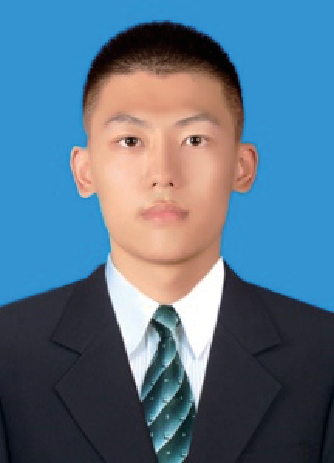}{Minglei Li}
received the B.Eng. degree in automation and the M.S. degree in control science and engineering from Qilu University of Technology (Shandong Academy of Sciences), Jinan, China, in 2017 and 2019, respectively. He is currently working toward the Ph.D. degree at the College of Control Science and Engineering, China University of Petroleum (East China). His current research interests include prognostics and health management (PHM) and deep learning.
\end{CCJNLbiography}

\end{document}